\documentclass[twocolumn,aps,preprintnumbers,prl,showpacs,amsfonts,amsmath,amssymb,superscriptaddress,floatfix]{revtex4}

\usepackage[dvips]{graphicx}
\usepackage[tight]{subfigure}

\allowdisplaybreaks[3]

\newcommand{\avg}[1]{\langle{#1}\rangle}

\newcommand{\affA}{%
    Department of Applied Physics, School of Engineering,
        The University of Tokyo,\\
    7-3-1 Hongo, Bunkyo-ku, Tokyo 113-8656, Japan}

\newcommand{\affB}{%
    CREST, Japan Science and Technology Agency,
    5 Sanbancho, Chiyoda-ku, Tokyo 102-0075, Japan}

\newcommand{\affC}{%
    Department of Physics, Technical University of Denmark, Building 309, 2800 Lyngby, Denmark}

\newcommand{\affD}{%
    Institut f\"{u}r Optik, Information und Photonik, Max-Planck Forschungsgruppe,
    Universit\"{a}t Erlangen-N\"{u}rnberg,
    G\"{u}nther-Scharowsky Stra{\ss}e 1, 91058 Erlangen, Germany}

\newcommand{\affE}{%
    OQI Group,
    Institute of Theoretical Physics I,
    Universit\"{a}t Erlangen-N\"{u}rnberg, Staudtstr. 7/B2,
    91058 Erlangen, Germany }

\begin{document}

\title{Demonstration of a quantum nondemolition sum gate}

\date{\today}

\author{Jun-ichi Yoshikawa}
\affiliation{\affA}
\affiliation{\affB}
\author{Yoshichika Miwa}
\affiliation{\affA}
\affiliation{\affB}
\author{Alexander Huck}
\affiliation{\affA}
\affiliation{\affC}
\affiliation{\affD}
\author{Ulrik L.\ Andersen}
\affiliation{\affC}
\affiliation{\affD}
\author{Peter van Loock}
\affiliation{\affD}
\affiliation{\affE}
\author{Akira Furusawa}
\affiliation{\affA}
\affiliation{\affB}


\begin{abstract}
The sum gate is the canonical two-mode gate for universal quantum computation based on continuous quantum variables.
It represents the natural analogue to a qubit C-NOT gate.
In addition, the continuous-variable gate describes a quantum nondemolition (QND) interaction
between the quadrature components of two light fields.
We experimentally demonstrate a QND sum gate, employing the scheme by R.\ Filip, P.\ Marek, and U.L.\ Andersen [\pra {\bf 71}, 042308 (2005)], solely based on offline squeezed states, homodyne measurements, and feedforward.
The results are verified by simultaneously satisfying the criteria for QND measurements in both conjugate quadratures.
\end{abstract}


\maketitle

The analogue of a two-qubit C-NOT gate, when continuous quantum variables
are considered, is the so-called sum gate. It represents the canonical
version of a two-mode entangling gate for universal quantum computation
in the regime of continuous variables \cite{Bartlett}.
When applied to two optical, bosonic modes,
as opposed to a simple beam splitter transformation, the sum gate
is even capable of entangling two modes each initially in
a coherent state, i.e., a close-to-classical state.

Apart from representing a universal two-mode gate,
the sum gate also describes a quantum nondemolition (QND) interaction.
The concept of a QND measurement has been known for almost 30 years.
Initially, it was proposed to allow for better accuracies in the
detection of gravitational waves~\cite{QND}.
A QND measurement is a projection measurement onto the basis
of a QND observable which is basically a constant of motion.
The QND measurement should preserve the measured observable,
but still gain sufficient information about its value;
the back action is confined to the conjugate observable.

Various demonstrations of QND or backaction evading measurements
have been reported~\cite{QND_experiment}.
The interest in the realization of a full QND {\it gate} grew only recently,
mainly in the context of continuous-variable (CV) quantum information
processing~\cite{SamPvLRMP}.
In particular, the QND sum gate is (up to local phase rotations)
the canonical entangling gate for building up Gaussian cluster
states~\cite{clusterZhang}, a sufficient resource for
universal quantum computation~\cite{clusterMenicucci}.
Other applications of the sum gate
are CV quantum error correction~\cite{QEC} and
CV coherent communication~\cite{coherent_communication}.

Here we report on the experimental demonstration of a full QND sum gate.
The gate leads to quantum correlations in {\it both} conjugate variables,
consistent with an entangled state, and allowing for a QND measurement of either variable with signal and probe interchanged.
While previous works focused on fulfilling the criteria for a QND measurement~\cite{QND_criteria} of one fixed variable, here we satisfy the QND criteria for two non-commuting observables, verifying entanglement at the same time.
As our implementation is very efficient and controllable, the current scheme can be used to process arbitrary optical quantum states, including fragile non-Gaussian states.
Similar to the measurement-based implementation of single-mode squeezing gates~\cite{Filip05.pra,Yoshikawa07.pra}, realization of the QND gate only requires two offline squeezed ancilla modes~\cite{Filip05.pra,Braunstein05.pra}.

Let us write the QND-gate Hamiltonian
as $\hat{H}_\text{QND}=\hat{x}_1\hat{p}_2$,
with a suitable choice of the absolute phase for each mode.
Here $\hat{x}/2$ and $\hat{p}/2$ are the real and imaginary parts of
each mode's annihilation operator, $\hat{a}=(\hat{x}+i\hat{p})/2$,
and the subscripts `1' and `2' denote two independent modes.
The ideal QND input-output relations then become,
\begin{align}
\hat{x}_1^\text{out}=&\hat{x}_1^\text{in},&
\hat{x}_2^\text{out}=&\hat{x}_2^\text{in}+G\hat{x}_1^\text{in}, \notag\\
\hat{p}_1^\text{out}=&\hat{p}_1^\text{in}-G\hat{p}_2^\text{in},&
\hat{p}_2^\text{out}=&\hat{p}_2^\text{in},
\label{eq:QND_ideal}
\end{align}
where $G$ is the gain of the interaction.

Through this ideal QND interaction,
the ``signal'' QND variable $\hat{x}_1$ ($\hat{p}_2$)
is preserved in the output state
and its value is added to the ``probe'' variable $\hat{x}_2$ ($\hat{p}_1$).
This allows for a QND measurement of either $\hat{x}_1$ or $\hat{p}_2$,
with a back action confined to the conjugate variable.
The usual criteria for QND measurements
(in the linearized, Gaussian regime) are~\cite{QND_criteria},
\begin{align}
1<T_\text{S}+T_\text{P}&\le 2, & V_{\text{S}|\text{P}} &< 1,& &
\label{eq:QND_criteria}
\end{align}
where $T_\text{S}$ and $T_\text{P}$ are the transfer coefficients
from signal input to signal output (``signal preservation'')
and from signal input to probe output (``information gain''),
respectively; $V_{\text{S}|\text{P}}$ is the conditional variance of
the signal output when the probe output is measured
(``quantum state preparation'').

\begin{figure}[t]
\centering
\includegraphics[clip,scale=0.65]{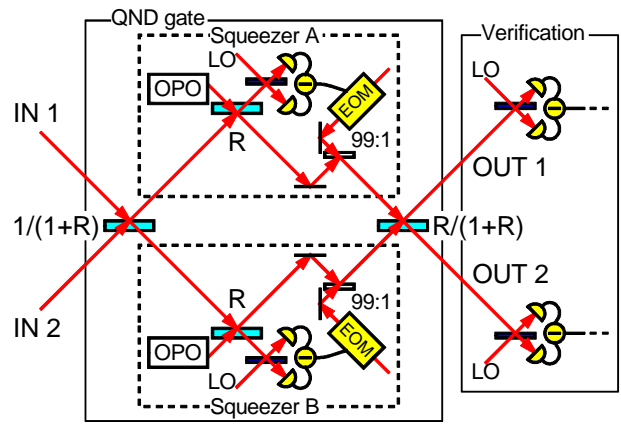}
\caption{(Colors online) Schematic of the experimental setup. The
parameter $R$ determines the reflectivities of the four
beam splitters, which are $1/(1+R),$ $R,$ $R,$ and $R/(1+R)$. We
employ optical parametric oscillators (OPO) to produce
ancilla squeezed vacuum modes, local oscillators (LO) for
homodyne detection, and electro-optic modulators (EOM) combined
with highly asymmetric beam splitters (99:1) for signal
displacement.} \label{fig:schematic}
\end{figure}

The implementation of the QND gate based on offline resources
is shown in Fig.~\ref{fig:schematic}.
The interaction gain $G$ in Eq.~\eqref{eq:QND_ideal} is related to
the reflectivities of the four beam splitters via one
free parameter $R$, with $G=\left(1/\sqrt{R}\right)-\sqrt{R}$,
taking arbitrary values for $0 < R \leq 1$.
The full scheme is described by the
input-output relations~\cite{Filip05.pra},
\begin{align}
\hat{x}_1^\text{out}=&\hat{x}_1^\text{in}-\sqrt{\frac{1-R}{1+R}}\,\hat{x}_\text{A}^{(0)}e^{-r_\text{A}},\\
\hat{x}_2^\text{out}=&\hat{x}_2^\text{in}+\frac{1-R}{\sqrt{R}}\hat{x}_1^\text{in}+\sqrt{R\,\frac{1-R}{1+R}}\,\hat{x}_\text{A}^{(0)}e^{-r_\text{A}}, \\
\hat{p}_1^\text{out}=&\hat{p}_1^\text{in}-\frac{1-R}{\sqrt{R}}\hat{p}_2^\text{in}+\sqrt{R\,\frac{1-R}{1+R}}\,\hat{p}_\text{B}^{(0)}e^{-r_\text{B}}, \\
\hat{p}_2^\text{out}=&\hat{p}_2^\text{in}+\sqrt{\frac{1-R}{1+R}}\,\hat{p}_\text{B}^{(0)}e^{-r_\text{B}},
\end{align}
where $\hat{x}_\text{A}^{(0)}e^{-r_\text{A}}$ and
$\hat{p}_\text{B}^{(0)}e^{-r_\text{B}}$ are the squeezed quadratures
of the ancilla modes, leading to some excess noise for finite
squeezing. The gate operation becomes ideal in the limit of infinite
squeezing ($r_\text{A},r_\text{B}\rightarrow\infty$).
Note that here precise control of active squeezing arising from an unstable process of parametric down conversion is not needed;
instead, the gate is completely controlled via passive optical devices.
For sufficiently large squeezing of the ancilla modes,
this transformation also allows for QND measurements.
Using variable beam splitters, we
experimentally realized two interaction gains, $G=1.0$ and $1.5$.
In particular, the unit gain interaction is significant for quantum
information processing. Nonetheless, we observe a better performance
in the QND measurements using higher gain.
We note that a non-unitary and single quadrature QND measurement based on squeezed vacuum and feedforward has been demonstrated in Ref.~\cite{Buchler02}.



{\it Experimental setup.}
A schematic of the experimental setup is illustrated in Fig.~\ref{fig:schematic}. It basically consists of a Mach-Zehnder interferometer with a single-mode squeezing gate in each arm. To implement fine-tunable and lossless squeezing operations, we use the measurement-induced squeezing approach proposed in Ref.~\cite{Filip05.pra}, experimentally implemented in Ref.~\cite{Yoshikawa07.pra} and illustrated inside the dashed boxes of Fig.~\ref{fig:schematic}.

We define the modes of the system to be residing at frequency sidebands of $1.25$~MHz relative to the optical carrier of the bright continuous wave light beam at a wavelenght of 860~nm from a Ti:sapphire laser.
The powers in each of the two input modes and the squeezed modes are 10~$\mu$W and 2~$\mu$W, respectively. These powers are considerably smaller than the powers (3~mW) of the local oscillators (LOs) used for homodyne detection.
All the interferences at the beam splitters are actively phase locked using modulation sidebands of 77~kHz, 106~kHz, and their beat in 29~kHz.
Subthreshold optical parametric oscillators (OPOs) generate the squeezed vacuum ancillas.
To control the beam splitting ratios of the four beam splitters in the squeezing operations and the Mach-Zehnder interferometer, they are composed of two polarizing beam splitters and a half wave plate~\cite{Yoshikawa07.pra}.

The OPOs are bow-tie shaped cavities of 500~mm in length, containing a periodically-poled KTiOPO$_4$ (PPKTP) crystal of 10~mm in length. The pump beams for the OPOs (with wavelengths of 430~nm and powers of about 100~mW) are the second harmonic of the output of the Ti:sapphire laser. The frequency doubling cavity (not shown in the figure) has the same configuration as the OPOs, but contains a KNbO$_3$ crystal. For details of a squeezed vacuum generation, see Ref.~\cite{Suzuki06.apl}.
Each OPO enables a squeezing degree of about $-5$~dB relative to the shot noise limit.

The outcomes of the homodyne detections in the QND gate are fed forward to the remaining part.
After low noise electric amplification, they drive an electro-optical modulator (EOM) traversed by an auxiliary beam with the power of 150~$\mu$W, which is subsequently mixed with the signal beam by a highly asymmetric beam splitter (99:1).

The QND scheme is characterized by measuring the two input modes as well as the two output modes using homodyne detection.
The detector's quantum efficiencies are higher than 99\%, the interference visibilities to the LOs are on average 98\%, and the dark noise of each homodyne detector is about 17~dB below the optical shot noise level produced by the local oscillator.
 We measure the propagation losses in each of the two main modes through the QND apparatus to be about 7\%.

\begin{figure}[tb]
\centering
\includegraphics[clip,scale=0.38]{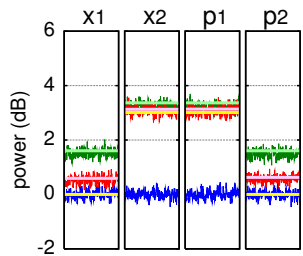}
 \caption{(Colors) Power spectra for input and output states corresponding to vacuum inputs and the interaction gain of $G=1.0$.
Shown are the experimental QND output variances (red) with their
theoretically calculated values (pink) compared to the results with
vacuum-state ancillas (green) and their theoretical values (light
green). The yellow lines refer to the theoretical results for infinite squeezing and the blue traces show the shot noise level corresponding to the variances of the input states.} \label{fig:vacuum}
\end{figure}

\begin{figure}[tb]
\centering
\subfigure[Amplitude in $x_1^\text{in}$.]{
\includegraphics[clip,scale=0.34]{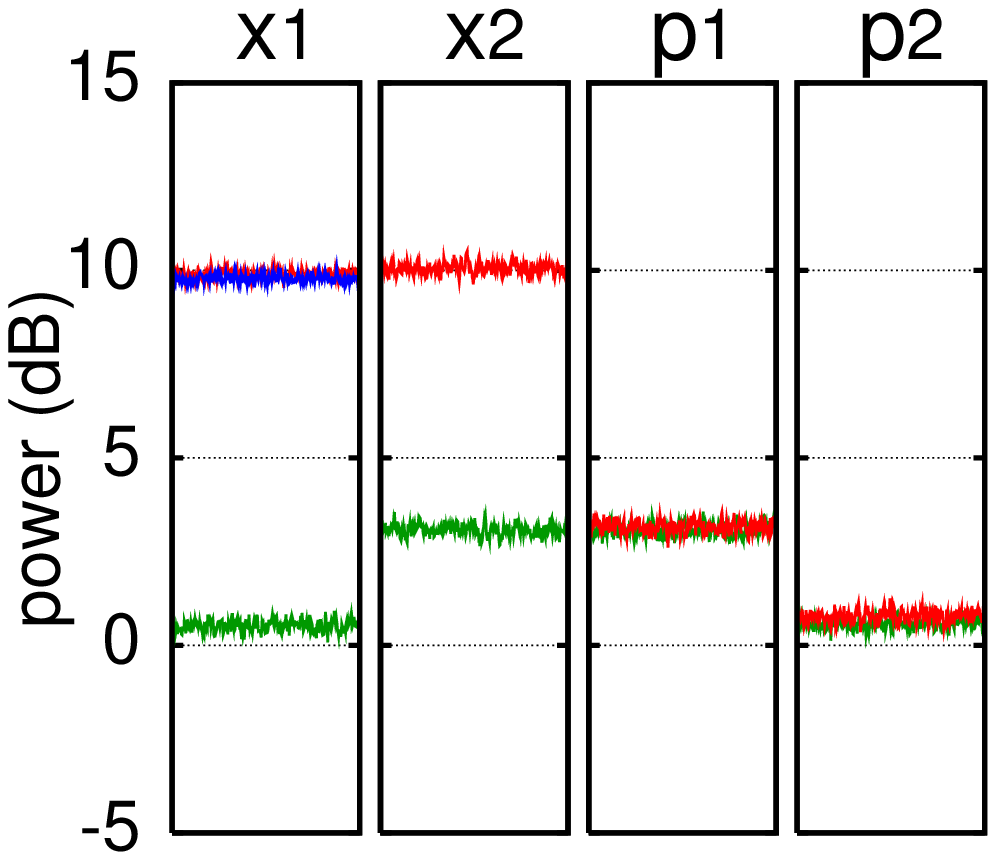}
}
\subfigure[Amplitude in $x_2^\text{in}$.]{
\includegraphics[clip,scale=0.34]{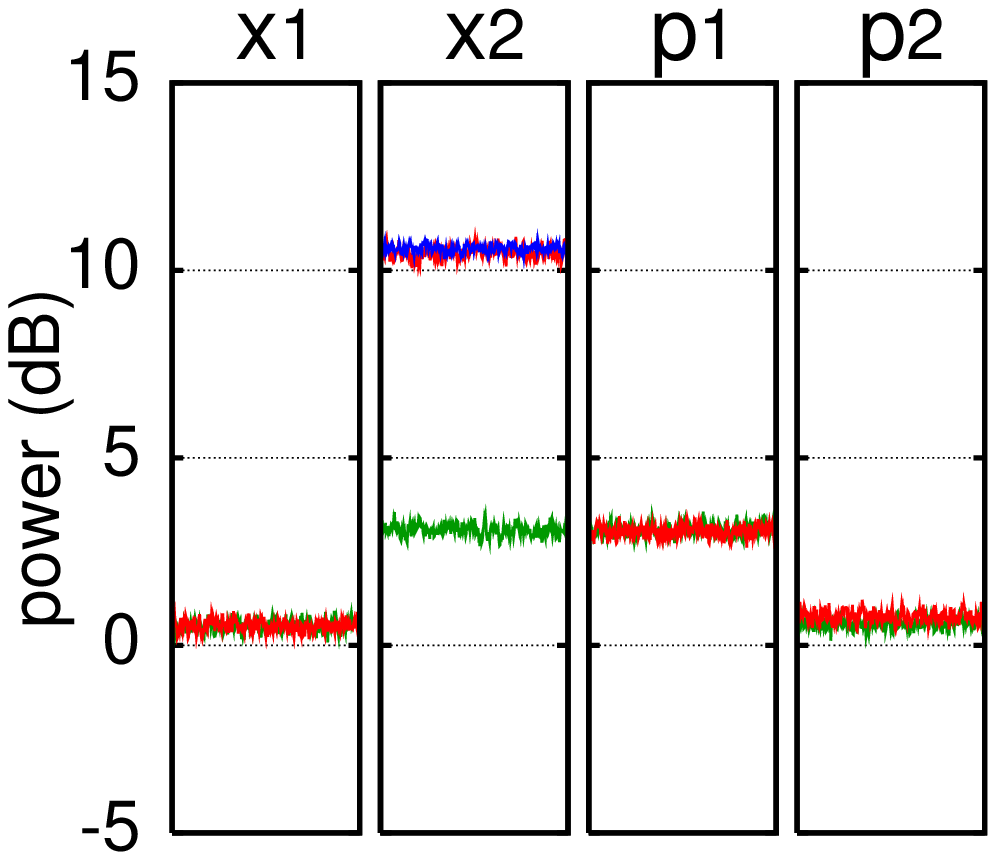}
}
\subfigure[Amplitude in $p_1^\text{in}$.]{
\includegraphics[clip,scale=0.34]{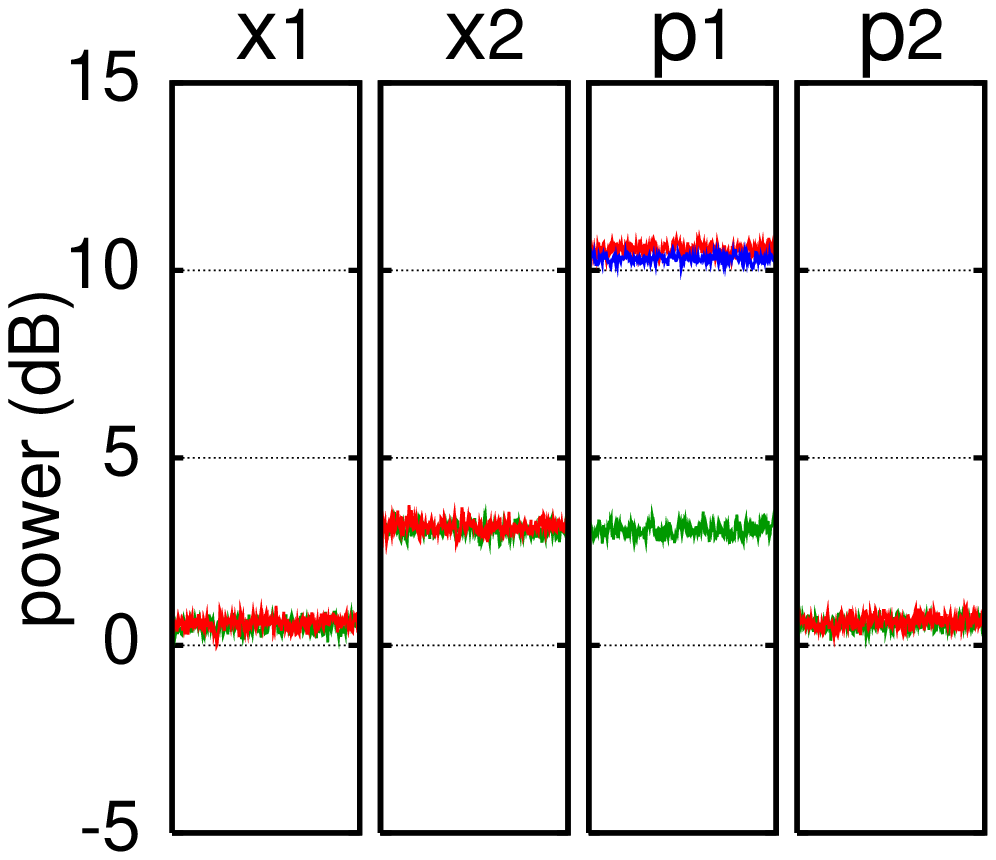}
}
\subfigure[Amplitude in $p_2^\text{in}$.]{
\includegraphics[clip,scale=0.34]{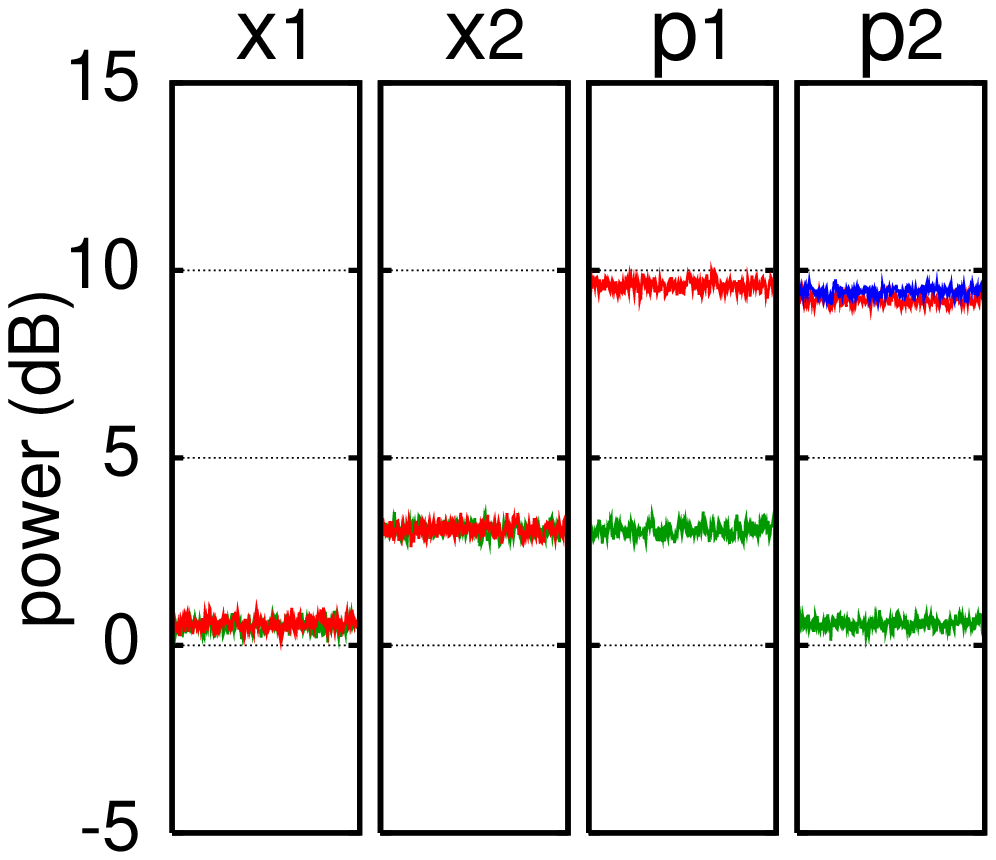}
} \caption{(Colors) Power spectra for the input and output states corresponding to four different input states for $G=1.0$. The excited input quadratures are shown in blue (non-excited input quadratures with variances equal to the shot noise level are not shown). The power spectra of the output quadratures are in red, and, for
comparison, we added the spectra corresponding to the measurements without squeezed ancilla states.} \label{fig:in-out}
\end{figure}

\begin{figure}[tb]
\centering
\subfigure[Setup of verification part.]{
\includegraphics[clip,scale=0.4]{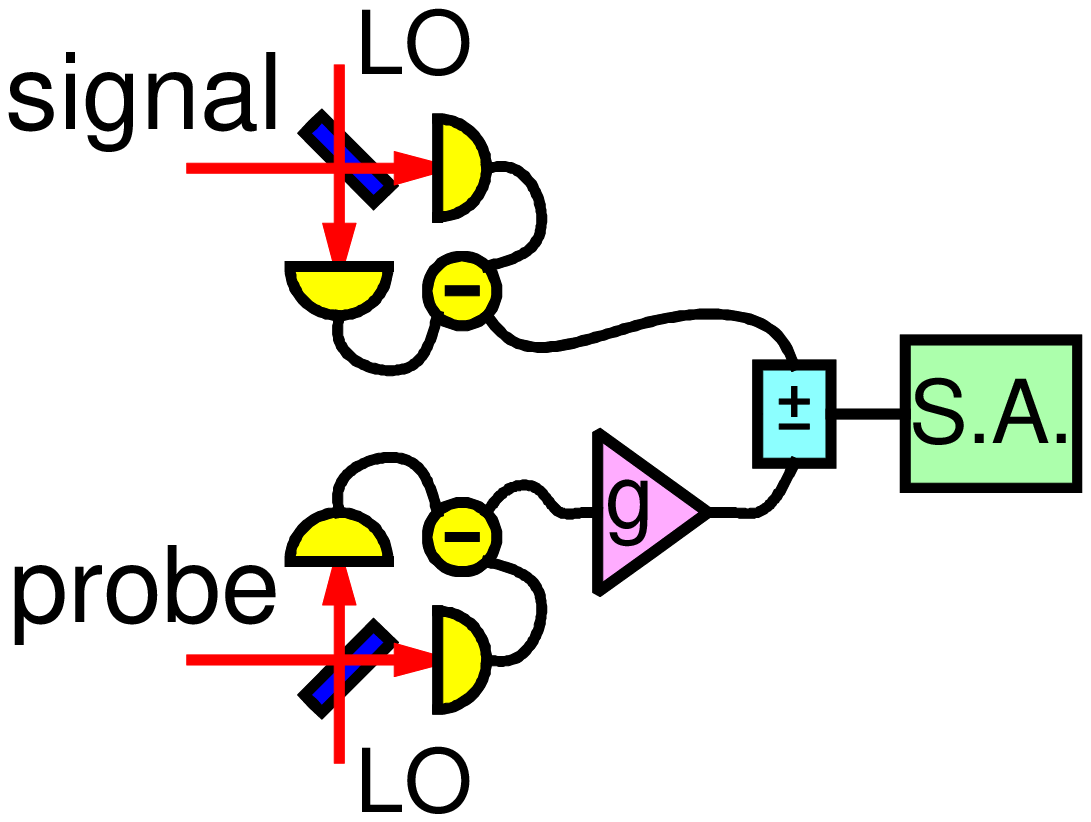}
}
\subfigure[Variances of ($\hat{x}_1^\text{out}-g\hat{x}_2^\text{out}$).]{
\includegraphics[clip,scale=0.35]{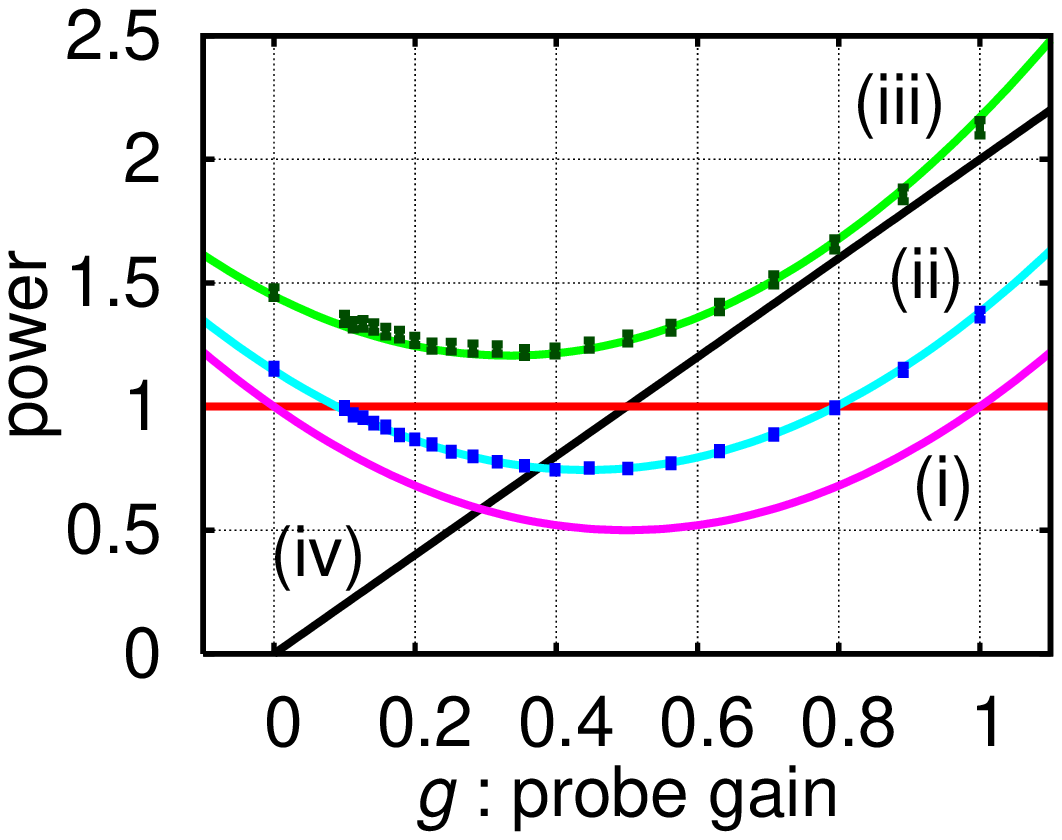}
}
\subfigure[Variances of ($\hat{p}_2^\text{out}+g\hat{p}_1^\text{out}$).]{
\includegraphics[clip,scale=0.35]{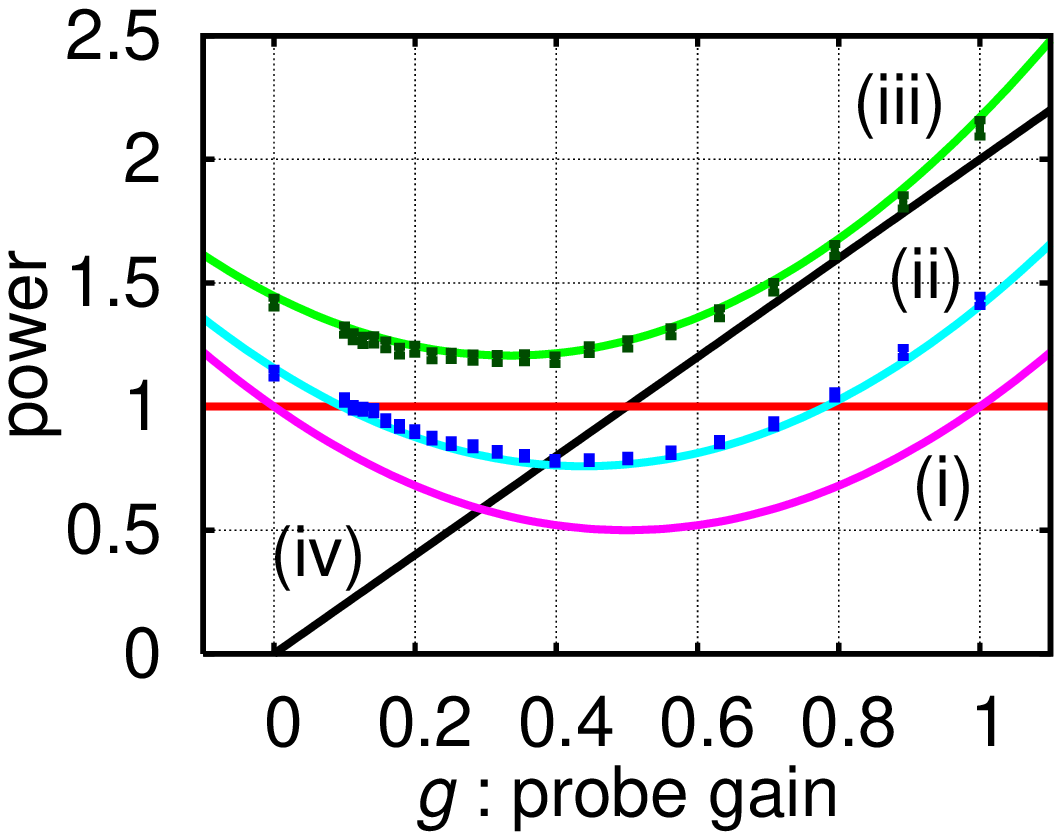}
} \caption{(Colors online) Differential quadrature power spectra of the two output states resulting in measurements of the conditional variances. (a)~Experimental setup for determining the conditional variance. The measured probe quadrature is rescaled with a variable gain
$g$, added (subtracted) ($\pm$) to (from) the signal mode detector
output, and analyzed with an electronic spectrum analyzer (S.A.).
Variances of
$\hat{x}_1^\text{out}-g\hat{x}_2^\text{out}$ and of
$\hat{p}_2^\text{out}+g\hat{p}_1^\text{out}$ are shown in (b) and (c), respectively:
theoretical prediction for an ideal QND interaction (i), a QND
interaction with a finite degree of squeezing of the ancilla modes
(ii), and with vacuum ancilla modes (iii). By entering the areas below the lines~(iv) 
entanglement is verified. The vertical axes are variances
normalized to the shot noise power of the signal variable.}
\label{fig:cond_var}
\end{figure}

{\it Experimental results.} The three measures in Eq.~\eqref{eq:QND_criteria} are used
to quantify the performance of our QND system.
To estimate them, we perform measurements of the second moments of the input fields and the output fields, employing a spectrum analyzer with a center frequency of 1.25~MHz, resolution and video bandwidths of 30~kHz and 300~Hz, respectively, a sweep time set to $0.1$~s and further averaging of 20 traces.
In Fig.~\ref{fig:vacuum}-\ref{fig:cond_var} the results for $G=1.0$ are shown as an example; in Table~\ref{table} the performance of the QND device is listed for both $G=1.0$ and $G=1.5$.

In the first series of measurements we determine the variances of conjugate quadratures of the output states when the input states are pure vacua.
The results corresponding to $G=1.0$ are presented in Fig.~\ref{fig:vacuum}.
The variances of the two input states are at the vacuum noise level as illustrated by the blue traces.
As a result of the QND interaction, in the ideal case, the noise of the signal variables ($\hat{x}_1$ and $\hat{p}_2$) is added to the probe variables ($\hat{x}_2$ and $\hat{p}_1$), while the signal variables are preserved.
The expected variances for the ideal performance is marked by the yellow lines and the actual measured variances of the output state is given
by the red traces. The deviation from the ideal performance is due to
the finite amount of squeezing for the ancillas.
For comparison, we also measure the variances of the output states when
no squeezing is used. This is shown by the green
traces. The expected variances, for finite
squeezing or without squeezing, are calculated and marked by pink and
light green lines, respectively.

In the second series of measurements,
 in order to test the universality of the QND gate, we replace the input vacuum states by a pair of coherent states.
 We generate the coherent amplitude in each quadrature of the two input modes by modulating the amplitude or phase of their carriers using an EOM operating at $1.25$~MHz. We investigate four different input states, each corresponding to a coherent excitation in, respectively, (a)~$x_1^\text{in}$, (b)~$x_2^\text{in}$, (c)~$p_1^\text{in}$ and (d)~$p_2^\text{in}$. The measurement results of the second moments of the input and output modes for $G=1.0$ are shown in Fig.~\ref{fig:in-out}. The excitations of the input modes are measured by setting the reflectivities of the four beam splitters to unity and blocking the auxiliary displacement beams in the feedforward construction. These measurements are illustrated by the blue traces (the non-excited quadratures are not shown because they are at the vacuum level, 0~dB). Traces in red are the second moments of the output modes. We observe that the amplitude of the input states is preserved in the same quadrature with almost unity gain. Furthermore, we clearly see the expected feature that the information in a signal variable, $\hat{x}_1^\text{in}$ or $\hat{p}_2^\text{in}$, is coupled into the probe variable $\hat{x}_2^\text{out}$ or $\hat{p}_1^\text{out}$ (see Fig.~\ref{fig:in-out}(a)~and~(d)), whereas the amplitude in the probe variables $\hat{x}_2^\text{in}$ and $\hat{p}_1^\text{in}$ does not couple to any of the other quadratures (see Fig.~\ref{fig:in-out}(b)~and~(c)). These results verify the interaction in Eq.~\eqref{eq:QND_ideal}. From these measurements we determine the transfer coefficients $T_\text{S}$ and $T_\text{P}$ using the method outlined in ref.~\cite{poizat94}. The results are summarised in table~\ref{table}.

Finally, we measure the conditional variance using the setup shown in Fig.~\ref{fig:cond_var}(a). The outcomes from one of the homodyne detectors are rescaled by a gain $g$, subtracted from (or added to) the outcomes of the other homodyne detector and subsequently directed to a spectrum analyzer. The resulting normalized noise powers are shown in Fig.~\ref{fig:cond_var}(b)~and~(c) as a function of the rescaling gain $g$. The minima of these plots correspond to the conditional variances for the various realizations: curve~(i) represents ideal performance, curve~(ii) is associated with our system with finitely squeezed ancillas, and curve~(iii) is the performance of the system without squeezed ancilla states. The parabolic curves are theoretical calculations, and the dots with vertical error bars along the curves~(ii)~and~(iii) are the experimental results.
The conditional variances, $V_{\text{S}|\text{P}}$, are collected in table~\ref{table}.

\begin{table}[tb]
\centering
\begin{tabular}{|c|c|c|c|c|}
 \hline
 G           &\multicolumn{2}{|c|}{1.0}      & \multicolumn{2}{|c|}{1.5} \\ \hline
 Quadrature  & $x$           & $p$           & $x$           & $p$  \\ \hline
 $T_\text{S}$       & $0.79\pm0.03$ & $0.71\pm0.03$ & $0.80\pm0.03$ & $0.71\pm0.03$ \\
 $T_\text{P}$       & $0.41\pm0.02$ & $0.39\pm0.02$ & $0.62\pm0.03$ & $0.56\pm0.02$ \\
 $T_\text{S}+T_\text{P}$   & $1.20\pm0.05$ & $1.10\pm0.05$ & $1.42\pm0.06$ & $1.27\pm0.05$ \\
 $V_{\text{S}|\text{P}}$   & $0.75\pm0.01$ & $0.78\pm0.01$ & $0.61\pm0.01$ & $0.63\pm0.01$ \\ \hline
\end{tabular}
\caption{Evaluation of the QND interaction. Shown are the
quadrature transfer coefficients $T_\text{S}$ and $T_\text{P}$ and the conditional
variance $V_{\text{S}|\text{P}}$ for two different gains G.
}
\label{table}
\end{table}

Our experiment demonstrates the realization of a canonical two-mode entangling gate.
From the output-output correlations in Fig.~\ref{fig:cond_var}, we verify entanglement between the two output modes. According to Duan and Simon~\cite{Duan00,Simon00}, a sufficient condition for an entangled state is,
\begin{equation}
\avg{(\hat{x}_1^\text{out}-g\hat{x}_2^\text{out})^2} + \avg{(\hat{p}_2^\text{out}+g\hat{p}_1^\text{out})^2} < 4 |g|,
\end{equation}
where $g$ is the rescaling gain. Thus, if the parabolic curves in Fig.~\ref{fig:cond_var}~(b) and~(c) go below the lines~(iv) {\it simultaneously} for both quadratures, the two output modes are entangled, which is the case for curves~(ii) with squeezed ancillas.
Note that the two-mode gate here has been applied to two coherent input states which, without the squeezed ancillas,
would not become entangled via {\it any} linear optical
transformation alone (see, e.g., curve~(iii)).\\

In conclusion, we have demonstrated and fully characterized a close-to-unitary quantum nondemolition sum gate using only linear optics and offline squeezed vacuum states.
The performance of the sum gate was quantified by applying the usual QND criteria to each conjugate quadrature; 
we found that the gate operates in the quantum regime, entangling even two input coherent states.
The future prospects of this demonstration are intriguing since the sum gate is an integral part of e.g. a one-way quantum computer based on continuous variables~\cite{clusterMenicucci} and quantum error correction protocols~\cite{QEC}.

This work was partly supported by SCF and GIA commissioned by the MEXT of Japan, and the Research Foundation for Opt-Science and Technology. ULA and AH acknowledge financial support from the EU under project No. 212008 (COMPAS) and the Lundbeck foundation.
PvL acknowledges support from the
Emmy Noether programme of the DFG in Germany.


\begin{thebibliography}{99}


\bibitem{Bartlett}
S.D.\ Bartlett {\it et al.},
\prl {\bf 88}, 097904 (2002).

\bibitem{QND}
C.M.\ Caves {\it et al.},
\rmp {\bf 52} 341 (1980).

\bibitem{QND_experiment}
For example,
Z.Y.\ Pereira {\it et al.},
\prl {\bf 72}, 214 (1994).

\bibitem{SamPvLRMP}
S.L.\ Braunstein and P.\ van Loock,
\rmp {\bf 77}, 513 (2005).

\bibitem{clusterZhang}
J.\ Zhang and S.L.\ Braunstein,
\pra {\bf 73}, 032318 (2006).

\bibitem{clusterMenicucci}
N.C.\ Menicucci {\it et al.},
\prl {\bf 97}, 110501 (2006).

\bibitem{QEC}
S.L.\ Braunstein, \prl {\bf 80}, 4084 (1998);
S.\ Lloyd and J.-J.E.\ Slotine, \prl {\bf 80}, 4088 (1998).

\bibitem{coherent_communication}
M.M.\ Wilde {\it et al.},
\pra {\bf 75}, 060303(R) (2007).



\bibitem{QND_criteria}
M.J.\ Holland {\it et al.},
\pra {\bf 42} 2995 (1990).


\bibitem{Filip05.pra}
R.\ Filip {\it et al.},
R.\ Filip, P.\ Marek, and U.L.\ Andersen,
Phys. Rev. A {\bf 71}, 042308 (2005).


\bibitem{Yoshikawa07.pra}
J.\ Yoshikawa {\it et al.},
\pra {\bf 76}, 060301(R) (2007).


\bibitem{Braunstein05.pra}
S.L.\ Braunstein, \pra {\bf 71}, 055801 (2005).



\bibitem{Buchler02}
B.C.\ Buchler {\it et al.},
\pra {\bf 65}, 011803(R) (2002).








\bibitem{Suzuki06.apl} S.\ Suzuki {\it et al.},
\apl {\bf 89}, 061116 (2006).

\bibitem{poizat94}
J.-Ph.\ Poizat {\it et al.},
Ann.\ Phys.\ (Paris) {\bf 19}, 265 (1994).




\bibitem{Duan00}
L.-M.\ Duan {\it et al.},
\prl {\bf 84}, 2722 (2000).

\bibitem{Simon00}
R.\ Simon, \prl {\bf 84}, 2726 (2000).




















\end{thebibliography}
\end{document}